# SimiSketch: Efficiently Estimating Similarity of streaming Multisets

Fenghao Dong*
Carnegie Mellon University
fenghaod@andrew.cmu.edu

Yang He†
Peking University
2000013166@stu.pku.edu.cn

Yutong Liang†
Peking University
lyt0112@stu.pku.edu.cn

Zirui Liu
Peking University
zirui.liu@pku.edu.cn

Yuhan Wu
Peking University
yuhan.wu@pku.edu.cn

Peiqing Chen
University of Maryland
pqchen99@umd.edu

Tong Yang‡
Peking University
yangtong@pku.edu.cn

## ABSTRACT
The challenge of estimating similarity between sets has been a significant concern in data science, finding diverse applications across various domains. However, previous approaches, such as MinHash [8], have predominantly centered around hashing techniques, which are well-suited for sets but less naturally adaptable to multisets, a common occurrence in scenarios like network streams and text data. Moreover, with the increasing prevalence of data arriving in streaming patterns, many existing methods struggle to handle cases where set items are presented in a continuous stream. Consequently, our focus in this paper is on the challenging scenario of multisets with item streams. To address this, we propose SimiSketch, a sketching algorithm designed to tackle this specific problem. The paper begins by presenting two simpler versions that employ intuitive sketches for similarity estimation. Subsequently, we formally introduce SimiSketch and leverage SALSA [7] to enhance accuracy. To validate our algorithms, we conduct extensive testing on synthetic datasets, real-world network traffic, and text articles. Our experiment shows that compared with the state-of-the-art, SimiSketch can improve the accuracy by up to 42 times, and increase the throughput by up to 360 times. The complete source code is open-sourced and available on GitHub for reference [3].

---

*The work is done at Peking University.

†Yang He and Yutong Liang contribute equally to the paper. They are listed alphabetically.

‡Tong Yang is the corresponding author.

---

## 1 INTRODUCTION
### 1.1 Background and Motivation

The estimation of similarity between sets has played a crucial role in data mining, finding diverse applications in areas such as data cleaning [6, 10, 34], network attack detection [12, 15, 19], and pattern matching [5, 14, 25, 29]. Previous research has predominantly focused on providing fast and memory-efficient estimates of set similarity through hashing approaches, including well-known techniques like MinHash [8, 17], Odd Sketch [22], MaxLogHash [30], DotHash [24], among others. However, due to the inherent nature of hashing, specifically the consistent mapping of the same item to a fixed value, it becomes challenging to determine the frequency of item occurrences. Consequently, these methods are unsuitable for cases involving multiset similarity estimation.

Furthermore, given the ubiquity of data streams in various domains like IoT data, network traffic, and database transactions, elements in sets may arrive in a streaming manner, necessitating rapid processing with only a single observation per item. Many existing approaches, such as [17, 22], are ill-suited for this streaming context, as highlighted in [30]. In contrast, sketching algorithms, a category of randomized algorithms proficient in estimating item frequency in data streams, have found extensive application in stream mining, network measurement, and database management systems [9, 11, 31]. Their adaptability makes them well-suited for addressing the challenge of estimating multisets' similarity in a streaming pattern, supporting flexible set operations like insert, delete, union, intersect, and sum. This work explores the application of sketching algorithms to meet the demands of estimating multisets' similarity in a streaming context.

One of the most widely used measures of set similarity is the Jaccard similarity, defined as $\frac{|A \cap B|}{|A \cup B|}$ for sets $A$ and $B$. In numerous scenarios, data can be conceptualized as items within a set, leading to the abstraction of the problem as the calculation of similarity between sets. A typical example is the representation of text documents as term vectors, where non-zero vector entries correspond to occurrences of words (or shingles). To identify similar text documents, Broder et al. [8] proposed calculating the Jaccard similarity of sets composed of words in each document. Accordingly, they designed and implemented MinHash to

efficiently estimate set similarity. However, a word (or shingle) may appear multiple times in a document, and for more precise results, it is necessary to include the word (or shingle) with its occurrence number. For instance, "a rose is a rose" would be represented as ("a", 1), ("rose", 1), ("is", 1), ("a", 2), ("rose", 2). Efficiently calculating the occurrence number requires the use of a hash table or sketching algorithms like count-min sketch [11]. However, a hash table demands significant memory and can be sluggish when conflicts occur. Count-Min sketch often overestimates the occurrence number, introducing errors in similarity estimation.

To address these challenges, we propose SimiSketch, a method that directly tackles the multiset problem without relying on hash tables or count-min sketch.

### 1.2 Our Proposed Solution

Towards the above design goal, this paper presents SimiSketch. Our high level idea is based on the divide and conquer strategy. First, we present two basic version of SimiSketch, which are intuitive, useful and easy to implement. The first one is the CM version of SimiSketch. Specifically, entirely using a Count-Min Sketch (or CM Sketch for short) [11]. We can consider the items mapped into a CM counter by a subset of the entire set. We found that the minimum of the corresponding CM counters is an over-estimation of the cardinality of the intersect, and the maximum of the corresponding CM counters is an under-estimation of the cardinality of the union, which will be analyzed in detail later. By adding up these minimums gives an over-estimation of cardinality of the intersect, and similarly adding up these maximums gives an under-estimation of cardinality of the union. Therefore it can give an over-estimation of the Jaccard similarity. The second one is the Count version of SimiSketch. The idea is that in CM version, there might be a lot of conflicting items in a counter, which mess up the estimation. Intuitively, only the items with large number of occurrence in both subsets play a significant role in the similarity. Using count sketch [9] by mapping items to $\{+1, -1\}$ randomly helps to reduce the impact of insignificant items, helping us focus on significant items better. So in this version we use a count sketch. Also considering the items mapped into a counter by a subset, we use an easy and intuitive way to estimate the similarity of each subset, and simply take the average of each subset to give the estimation. The experiment results show that this easy and intuitive approach can give pretty accurate estimation, especially in small memory, which indicates the estimation using count sketch is somewhat promising. Inspired by the findings, we present SimiSketch, a more delicate and accurate one for estimating multiset similarity. The idea is that the similarity of the set can be expressed by a linear combination of the similarity of each subsets, so we introduce a CM counter to each counter in the count sketch to estimate the coefficients of the linear combination. Using the subset similarity estimated by count sketch counter and the coefficient estimated by CM counter gives the estimation. Furthermore, we use SALSA [7] to organize the count sketch in the form of circle array, and merge adjacent counters clockwise when overflow occurs. This can help to increase the accuracy of SimiSketch a lot.

The remainder of this paper is organized as follows: We provide the formal definition of the problem in Section 2.1, followed by an overview of existing solutions in section 2.2 and brief introductions to sketching algorithms in Section 2.3. Subsequently, we present the two basic versions in Section 3.1 and 3.2, introduce SimiSketch in Section 3.3, and discuss optimization using SALSA in Section 3.4. Finally, we present the experimental results in Section 4.

## 2 BACKGROUND AND RELATED WORK

### 2.1 Problem Statement

In this paper we consider multiset, i.e. sets where items can have multiple occurrences. Suppose $A$ is a multiset, we define its multiplicity function by $m_A$, specifically,

$$m_A(x) = \begin{cases} \text{multiplicity of x in A, if } x \in A \\ 0, \text{if } x \notin A \end{cases}$$

Define the support of $A$ by $Supp(A) := \{x : m_A(x) > 0\}$, which is a set. The cardinality of A is $|A| := \sum_{x \in Supp(A)} m_A(x)$. Given a set $S$, note that the multisets supported on $S$ are 1-1 corresponding to the mappings from $S$ to $\mathbb{Z}_{>0}$, mapping items in $S$ to its number of occurrence in the multiset. Specifically, mapping $\Phi_S$

$$\Phi_S : \{\text{multiset supported on } S\} \longrightarrow \{\text{mappings: } S \to \mathbb{Z}_{>0}\}$$
$$A \longmapsto \Phi_S(A) = \{x \mapsto m_A(x)\}$$

is an one-to-one mapping. Therefore we can define a multiset by its support $S$ and its image under mapping $\Phi_S$ defined above (or equivalently, $m_A$, since $\Phi_S(A) = m_A|_S$).

There are four basic operations on multisets. Suppose $A, B$ are multisets:

- **Sum** $A \oplus B$: support is $Supp(A) \cup Supp(B)$, multiplicity function is $m_{A \oplus B}(x) = m_A(x) + m_B(x)$.
- **Union** $A \cup B$: support is $Supp(A) \cup Supp(B)$, multiplicity function is $m_{A \cup B}(x) = \max(m_A(x), m_B(x))$.
- **Intersect** $A \cap B$: support is $Supp(A) \cap Supp(B)$, multiplicity function is $m_{A \cap B}(x) = \min(m_A(x), m_B(x))$.
- **Difference** $A \setminus B$: support $\subseteq Supp(A)$, multiplicity function is $m_{A \setminus B}(x) = \max(m_A(x) - m_B(x), 0)$.

As for similarity metric, we focus on Jaccard similarity measure, one of the most popular ones, as lots of previous works do [8, 17, 22, 30]. The Jaccard similarity of multiset $A$ and $B$ is defined as below:

$$J_{A,B} := \frac{|A \cap B|}{|A \cup B|}$$

Note that it looks the same as the set case, but the intersect, union and cardinality are in the notion of multiset.

### 2.2 Existing Solutions

In this section, we show some of the typical solutions regarding the set similarity problem. Some of them are classical solutions, like MinHash [8] and HyperLogLog [13], others are recent work published in KDD, like MaxLogHash [30] and DotHash [24]. Note that number of previous work doesn't fit in the streaming context, like b-bit MinHash, Odd Sketch, as analyzed in [30], we only focus on existing solutions which work for streaming sets.

*2.2.1 MinHash.* Published in 1998 by Broder et al., MinHash [8] is a very classical algorithm in estimating Jaccard similarity between sets. At high level, MinHash is based on the idea of sampling. Intuitively, for set $A$ and $B$, when we sample a subset of each of them, the portion of intersection of the two subsets is roughly the Jaccard similarity between $A$ and $B$. Formally, the MinHash algorithm uses $k$ hash functions, denote by $h_i$, $i = 1, 2, ..., k$. Each $h_i$ maps items in $A \cap B$ to integers in $[0, M]$ randomly. Broder et al. observed that

$$\mathbb{E}\left(\frac{1}{k}\sum_{i=1}^{k}\mathbb{1}\left\{\min_{x \in A} h_i(x) = \min_{y \in B} h_i(y)\right\}\right) = \frac{|A \cap B|}{|A \cup B|} = J_{A,B}$$

So they take

$$\hat{J}_{A,B} = \frac{1}{k}\sum_{i=1}^{k}\mathbb{1}\left\{\min_{x \in A} h_i(x) = \min_{y \in B} h_i(y)\right\}$$

as an estimator of $J_{A,B}$, which is unbiased.

*2.2.2 HyperLogLog.* Published by Flajolet et al., HyperLogLog [13] is a classical algorithm in estimating set cardinality. For each item in set $A$, it needs to compute an $N$ bit hash value of it. Take $M < N$, maintain an array of $L = 2^M$ counters $a_0, a_1, ..., a_{L-1}$, initially all zero. For each item $x \in A$, define the hash value of it by $h(x)$, which is of $N$ bits. Use its initial $M$ bits $h(x)[0:M-1]$ to select a counter from the counter array, specifically $a_{h(x)[0:M-1]}$. Define function $\rho(x)$ by the position of the leftmost 1-bit of x. For example, $\rho(1...) = 1, \rho(0001...) = 4, \rho(0^k) = k + 1$. The update procedure for each $x \in A$ is:

$$a_{h(x)[0:M-1]} = \max\left(a_{h(x)[0:M-1]}, \rho(h(x)[M:N-1])\right)$$

After updating these counters, compute estimate $E$ as below:

$$E = \alpha_L L^2 \cdot \left(\sum_{i=0}^{L-1} 2^{-a_i}\right)^{-1}$$

where $\alpha_L = \frac{0.7213}{1+\frac{1.079}{L}}$ for $L \geq 128$. When $2.5L < E \leq \frac{1}{30}2^{32}$, $E$ is the estimator to the cardinality of set $A$. Consider set $A$ and $B$, if we use the same parameter setting and the same hash functions $h_i$ for HyperLogLog, resulted in counter arrays $a_0, a_1, ..., a_{L-1}$ and $b_0, b_1, ..., b_{L-1}$ respectively. Therefore we can get the estimation of $|A|$ and $|B|$. It's easy to see that if we'd use the same parameter setting and hash functions for set $A \cup B$, resulted in counter arrays $c_0, c_1, ..., c_{L-1}$, then $c_i = \max(a_i, b_i)$ will hold for $\forall i = 0, 1, ..., L - 1$. Therefore, $|A \cup B|$ can be derived directly from counter arrays $\vec{a}$ and $\vec{b}$. Since $J_{A,B} = \frac{|A \cap B|}{|A \cup B|} = \frac{|A|+|B|-|A \cup B|}{|A \cup B|}$, we can use HyperLogLog to estimate set similarity.

*2.2.3 MaxLogHash.* MaxLogHash is recently published by Wang et al. [30]. At high level, it's also based on the idea of sampling, pretty similar to that in MinHash. For each set $S^{(j)}$ ($j = 0, 1$), it maintains an array $a_0^{(j)}, a_1^{(j)}, ..., a_{k-1}^{(j)}$ of length $k$ indicates the maximum log hash value, and an array of bool value $b_0^{(j)}, b_1^{(j)}, ..., b_{k-1}^{(j)}$, where $b_i^{(j)}$ indicates whether the maximum hash value in $a_i^{(j)}$ is unique. Initially set $a_i^{(j)} = -1$ and $b_i^{(j)} = 1$ for $\forall i = 0, 1, ..., k - 1. j = 0, 1$. Pick $k$ independent hash functions $h_0, h_1, ..., h_{k-1}$ mapping items in the set to $(0, 1)$ uniformly. For each coming item $x^{(j)} \in A^{(j)}$, denote $l_i^{(j)} := \left\lfloor -\log(h_i(x^{(j)}))\right\rfloor$, the updating procedure is:

- if $l_i^{(j)} < a_i$, do nothing.
- if $l_i^{(j)} = a_i$, set $b_i = 0$.
- if $l_i^{(j)} > a_i$, set $a_i = l_i^{(j)}$, $b_i = 1$.

After inserting all items from $S^{(0)}$ and $S^{(1)}$, compute the following values for $\forall i = 0, 1, ..., k - 1$:

- $\chi_i = \mathbb{1}\left\{a_i^{(0)} \neq a_i^{(1)}\right\}$.
- $\phi_i = \sum_{j=0}^{1} b_i^{(j)} \cdot \mathbb{1}\left\{a_i^{(j)} > a_i^{(1-j)}\right\}$.
- $\delta_i = \chi_i \cdot \phi_i$

The estimator of $J_{S^{(0)},S^{(1)}}$ is

$$\hat{J}_{S^{(0)},S^{(1)}} = 1 - \frac{1}{k \cdot \alpha_{|S^{(0)} \cup S^{(1)}|}}\sum_{i=0}^{k-1}\delta_i$$

where $\alpha_n \approx 0.7213$ for $n \geq 2$.

*2.2.4 DotHash.* Published by Nunes et al., DotHash is a novel algorithm based on the fact that random vectors in high dimensional space is almost orthogonal with high probability. Specifically, take mapping $\phi$ maps items to unit vectors in $\mathbb{R}^d$ randomly. For set $A$ and $B$, calculate $a = \sum_{x \in A}\phi(x)$ and $b = \sum_{x \in B}\phi(x)$. Then $a \cdot b$ is an unbiased estimator for $|A \cap B|$, since $J_{A,B} = \frac{|A \cap B|}{|A \cup B|} = \frac{|A \cap B|}{|A|+|B|-|A \cap B|}$, we can use it to estimate $J_{A,B}$.

## 2.3 Sketching Algorithms

Sketching algorithm is a kind of streaming algorithm solving problems like frequency estimation, finding top-K items, etc. Basically, it uses hash functions to map items to a slot, then do some updates like increasing the counter, store the key and value, etc. Its main differences from hash tables are as below:

- Most sketching algorithms can guarantee $O(1)$ insertion time, while hash tables can't.
- Most sketching algorithms use a fix amount of memory, while some hash tables don't, like separate chaining.
- Sketching algorithms can only give an estimation to the frequency. But for hash tables, if there's enough memory to store a given item, it can always report the precise frequency of the item.

Also note that in order to make hash tables reach high performance, we need to allocate lots of memory to it, usually several times of that just enough to store all the distinct items. If we allocate pretty small memory, less than that just capable to store all the distinct items, it will only record some of the items, based on the first occurrence time. For items show up early, it will accurately report its frequency. However, for items show up late, it won't record it, and can not give any meaningful estimation. But for sketching algorithms, it will always try to give more accurate estimation to significant items (i.e. items appears most frequently), no matter how much memory allocated. In practice, stream processing tasks like network measurement often have following properties: i) They need high throughput. ii) We don't know and can not predict how many distinct items in total. Therefore, sketching algorithms are ideal for these kinds of tasks, rather than hash tables.

Sketching algorithms have been a hot research topic, with lots of work proposing algorithmic design [16, 18, 27, 28, 32], with other work focused on its estimation error [21, 26] and configuration [4, 20, 33]. In this section we introduce two classical sketching algorithm, count-min (CM) sketch [11] and count sketch [9].

*2.3.1 Count-Min (CM) Sketch.* Count-min sketch [11] uses a data structure of $k$ counter arrays, each of length $l$. Initially all of these counters are zero. There's also $k$ independent hash functions, mapping each item in a stream randomly and uniformly to one of the $l$ counters in an array. Denote the counters by $a[i][j]$, meaning the $j$-th counter in the $i$-th counter array, and denote the hash functions by $h_i$, where $i = 0, 1, 2, ..., k - 1$. For each coming item $x$ in the stream, its update process is

$$a[i][h_i(x)] = a[i][h_i(x)] + 1$$

for each $i = 0, 1, 2, ..., k - 1$, which is shown in figure 1. After inserting all the items, if we want to query the frequency of $x$, it's easy to see that each $a[i][h_i(x)]$ is an over-estimation of it. So the estimator is $\min_{i=0,1,...,k-1} a[i][h_i(x)]$, which is also an over-estimation.

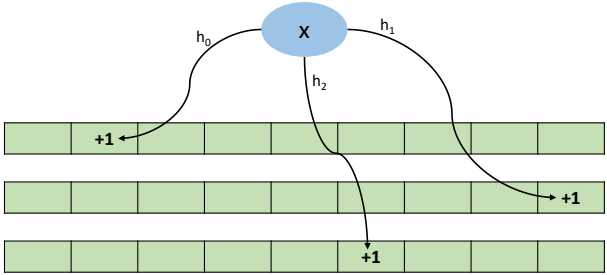

Figure 1: Count-Min Sketch

*2.3.2 Count Sketch.* Count sketch [9] use the same data structure as the count-min sketch, it uses counter arrays and independent hash functions mapping items into a slot in a row, as we introduced before. Besides of these, it also needs $k$ independent hash functions $s_i$ mapping items randomly and uniformly into $\{+1, -1\}$. Initially all the counters $a[i][j]$ are zero. For each coming item $x$ in the stream, its update process is

$$a[i][h_i(x)] = a[i][h_i(x)] + s_i(x)$$

After inserting all the items, we can get $k$ independent estimators $a[i][h_i(x)] \cdot s_i(x)$ to the frequency of $x$, and each of them is unbiased. So we take the median of these $k$ estimators to give the estimation.

## 3 THE SIMISKETCH

In this section we formally introduce SimiSketch. First we present two basic versions of SimiSketch, the CM version and the count version, which are based on the observation that using the same data structure as Count-Min sketch and count sketch can in fact give an estimation to the set similarity. Then based on the two basic versions, we introduce our formal version of SimiSketch, which is a combination of them. Afterwards, we will introduce the SALSA technique [7] and how to combine it with SimiSketch to increase the accuracy.

### 3.1 Simple version #1: The CM Version

We observe that using the similar data structure as count-min sketch can actually give estimation to Jaccard similarity of (streaming) multisets. Specifically, for multisets $A$ and $B$, we need two sets of $k$ counter arrays with length $l$, denote these counters by $a[i][j]$ and $b[i][j]$, where $i = 0, 1, ..., k - 1$ and $j = 0, 1, 2, ..., l - 1$. Also, there are $k$ independent hash functions $h_0, h_1, ..., h_{k-1}$ mapping items in $A \cup B$ randomly and uniformly into $\{0, 1, 2, ..., l - 1\}$. for each coming item $x \in A$ and $y \in B$, the update process is:

$$a[i][h_i(x)] = a[i][h_i(x)] + 1$$
$$b[i][h_i(y)] = b[i][h_i(y)] + 1$$

for each $i = 0, 1, ..., l - 1$. Denote the items mapped into $a[i][j]$ and $b[i][j]$ by $A_{ij}$ and $B_{ij}$, which are subsets of $A$ and $B$ respectively. We made several observations below:

LEMMA 3.1. $\min\{a[i][j], b[i][j]\} \geq |A_{ij} \cap B_{ij}|$ holds for $\forall i \in \{0, 1, 2, ..., k-1\}$ and $\forall j \in \{0, 1, 2, ..., l-1\}$.

PROOF. Denote $S_{ij} := Supp(A_{ij} \cup B_{ij})$, the multiplicity function of $A$ and $B$ by $m_A$ and $m_B$ respectively. Then it's easy to see:

$$a[i][j] = \sum_{x \in S_{ij}} m_A(x)$$
$$b[i][j] = \sum_{x \in S_{ij}} m_B(x)$$

Therefore

$$|A_{ij} \cap B_{ij}| = \sum_{x \in S_{ij}} \min\{m_A(x), m_B(x)\}$$

$$\leq \min\left\{\sum_{x \in S_{ij}} m_A(x), \sum_{x \in S_{ij}} m_B(x)\right\} = \min\{a[i][j], b[i][j]\}$$

which gives the proof. □

LEMMA 3.2. $\max\{a[i][j], b[i][j]\} \leq |A_{ij} \cup B_{ij}|$ holds for $\forall i \in \{0, 1, 2, ..., k-1\}$ and $\forall j \in \{0, 1, 2, ..., l-1\}$.

PROOF. We use the same notation as the previous proof. Similarly,

$$|A_{ij} \cap B_{ij}| = \sum_{x \in S_{ij}} \max\{m_A(x), m_B(x)\}$$

$$\geq \max\left\{\sum_{x \in S_{ij}} m_A(x), \sum_{x \in S_{ij}} m_B(x)\right\} = \max\{a[i][j], b[i][j]\}$$

which gives the proof. □

LEMMA 3.3.

$$\sum_{j=0}^{l-1} \min\{a[i][j], b[i][j]\} \geq |A \cap B|$$

$$\sum_{j=0}^{l-1} \max\{a[i][j], b[i][j]\} \leq |A \cup B|$$

holds for $\forall i = 0, 1, 2, ..., k - 1$.

Proof. Due to the property of hash functions $h_i$, we can see that $A_{ij_1} \cap A_{ij_2} = \emptyset$, $B_{ij_1} \cap B_{ij_2} = \emptyset$, $A_{ij_1} \cap B_{ij_2} = \emptyset$ holds $\forall i, \forall j_1 \neq j_2$. Therefore,

$$\begin{cases} A \cap B = \sum_{j=1}^{l-1} A_{ij} \cap B_{ij} \\ A \cup B = \sum_{j=1}^{l-1} A_{ij} \cup B_{ij} \end{cases} \Longrightarrow \begin{cases} |A \cap B| = \sum_{j=1}^{l-1} |A_{ij} \cap B_{ij}| \\ |A \cup B| = \sum_{j=1}^{l-1} |A_{ij} \cup B_{ij}| \end{cases}$$

Combining with Lemma 3.1 and 3.2 gives the proof. □

According to Lemma 3.3, $\frac{\sum_{j=0}^{l-1} \min\{a[i][j], b[i][j]\}}{\sum_{j=0}^{l-1} \max\{a[i][j], b[i][j]\}}$ is an over-estimation to $J_{AB}$ for $\forall i = 0, 1, ..., k - 1$. So we take our estimator as

$$\hat{J}_{AB} = \min_{i=0,1,...,k-1} \left\{ \frac{\sum_{j=0}^{l-1} \min\{a[i][j], b[i][j]\}}{\sum_{j=0}^{l-1} \max\{a[i][j], b[i][j]\}} \right\}$$

which is also an over-estimation.

In fact, this estimator can also be derived in an divide and conquer point of view. since

$$J_{AB} = \frac{|A \cap B|}{|A \cup B|} = \frac{\sum_{j=0}^{l-1} |A_{ij} \cap B_{ij}|}{|A \cup B|} = \sum_{j=0}^{l-1} \frac{|A_{ij} \cup B_{ij}|}{|A \cup B|} \cdot \frac{|A_{ij} \cap B_{ij}|}{|A_{ij} \cup B_{ij}|} \quad (1)$$

$$= \sum_{j=0}^{l-1} \frac{|A_{ij} \cup B_{ij}|}{|A \cup B|} \cdot J_{A_{ij}B_{ij}} \quad (2)$$

Therefore, in order to estimate $J_{AB}$, we can device $A$ and $B$ to subsets $A = \sum_{j=0}^{l-1} A_{ij}$ and $B = \sum_{j=0}^{l-1} B_{ij}$. After estimating each $J_{A_{ij}B_{ij}}$, combining them together gives the estimation to $J_{AB}$. According to Lemma 3.1 and 3.2, $\frac{\min\{a[i][j], b[i][j]\}}{\max\{a[i][j], b[i][j]\}}$ is an estimator of $J_{A_{ij}B_{ij}}$. As for combination coefficient, $\max\{a[i][j], b[i][j]\}$ is an estimator of $|A_{ij} \cup B_{ij}|$, $\sum_{k=0}^{l-1} \max\{a[i][k], b[i][k]\}$ is an estimator of $|A \cup B|$. As the result, our estimator can be written as:

$$\sum_{j=0}^{l-1} \frac{\max\{a[i][j], b[i][j]\}}{\sum_{k=0}^{l-1} \max\{a[i][k], b[i][k]\}} \cdot \frac{\min\{a[i][j], b[i][j]\}}{\max\{a[i][j], b[i][j]\}}$$

$$= \frac{\sum_{j=0}^{l-1} \min\{a[i][j], b[i][j]\}}{\sum_{j=0}^{l-1} \max\{a[i][j], b[i][j]\}}$$

Which is the same as what we previously got.

**Rationale:** The 80-20 rule is very common in data streams, specifically only a small number of distinct items contribute to a very large fraction of the entire stream, since they show up many times. We define $\varepsilon$-support of multiset $A$ by a set $A' \subset Supp(A)$, where $\sum_{x \in A'} m_A(x) \geq (1-\varepsilon)|A|$, denoted by $Supp_\varepsilon(A)$. Define the $\varepsilon$-subset of multiset $A$ by a subset of $A$ supported on $Supp_\varepsilon(A)$ with multiplicity function $m(x) = m_A(x) \cdot \mathbb{1}\{x \in Supp_\varepsilon(A)\}$. Note that $\varepsilon$-support and $\varepsilon$-subset are not unique. Intuitively, if the 80-20 rule applies to $A$, we'd expect $|Supp_\varepsilon(A)|$ much smaller than $|Supp(A)|$ for a relative small $\varepsilon$.

Intuitively, we'd call the items in $Supp_\varepsilon(A)$ significant, since they contributes a lot to $|A|$. The following lemma shows that the significant items also contributes a lot to $J_{AB}$.

Lemma 3.4. Suppose $A'$ and $B'$ are the $\varepsilon$-subset of multiset $A$ and $B$, then $|J_{AB} - J_{A'B'}| < 2\varepsilon$ holds for $\forall \varepsilon \in (0, 1)$.

Proof. Denote $A'' := A \setminus A'$, $B'' := B \setminus B'$. By the defination of $\varepsilon$-subset, $A' \cap A'' = \emptyset$, $B'' \cap B = \emptyset$. We have

$$|A \cap B| = |A' \cap B| + |A'' \cap B| = |A' \cap B'| + |A' \cap B''| + |A'' \cap B|$$
$$\leq |A' \cap B'| + \varepsilon|A| + \varepsilon|B|$$

$$|A \cup B| = |A| + |B| - |A \cap B| \leq \frac{1}{1-\varepsilon}(|A'| + |B'|) - |A' \cap B'|$$
$$= |A' \cup B'| + \frac{\varepsilon}{1-\varepsilon}(|A'| + |B'|)$$

Therefore,

$$\frac{|A \cap B|}{|A \cup B|} \leq \frac{|A' \cap B'|}{|A \cup B|} + \varepsilon \frac{|A| + |B|}{|A \cup B|}$$
$$\leq \frac{|A' \cap B'|}{|A' \cup B'|} + \varepsilon\left(1 + \frac{|A \cap B|}{|A \cup B|}\right)$$
$$\leq \frac{|A' \cap B'|}{|A' \cup B'|} + 2\varepsilon$$

On the other hand,

$$\frac{|A \cap B|}{|A \cup B|} \geq \frac{|A' \cap B'|}{|A' \cup B'| + \frac{\varepsilon}{1-\varepsilon}(|A'| + |B'|)}$$
$$= J_{A'B'} \cdot \frac{1}{1 + \frac{\varepsilon}{1-\varepsilon} \cdot \frac{|A'|+|B'|}{|A' \cup B'|}}$$
$$\geq \frac{1-\varepsilon}{1+\varepsilon} J_{A'B'} \geq J_{A'B'} - 2\varepsilon$$

which gives the proof. □

The error of count-min sketch comes from collision. Intuitively, if there's no collisions in our counter array, our estimation to $J_{AB}$ will be accurate without error. However, in practice there's always some collisions which brings the error. But since insignificant items don't show up often, they don't have big effect when collision occurs. So when significant items don't collide with each other, the resting insignificant items won't have much effect, therefore our estimation will be relatively accurate. Based on this idea, we can derive the following error bound:

Theorem 3.5. For $\forall \varepsilon \in (0, 1)$, denote $k := |Supp_\varepsilon(A) \cup Supp_\varepsilon(B)|$. For $\forall i = 0, 1, ..., k-1$, denote $\hat{J}_{AB}^{(i)} := \frac{\sum_{j=0}^{l-1} \min\{a[i][j], b[i][j]\}}{\sum_{j=0}^{l-1} \max\{a[i][j], b[i][j]\}}$ by the estimation given by the i-th row of CM version. Then $|J_{AB} - \hat{J}_{AB}^{(i)}| < 2\varepsilon + \frac{2\varepsilon}{1-\varepsilon}$ holds with probability $p \geq \frac{l!}{(l-k)!l^k}$.

Proof. Denote the $\varepsilon$-subset of $A$ and $B$ by $A'$ and $B'$. Since $k$ items in $Supp_\varepsilon(A) \cup Supp_\varepsilon(B)$ are mapped randomly and uniformly into a CM array of length $l$, the probability of they don't conflict with each other is $\frac{l(l-1)(l-2)...(l-k+1)}{l^k} = \frac{l!}{(l-k)!l^k}$. We only consider the case where this happens.

When these $k$ items don't conflict with each other, we can see that

$$|A' \cap B'| \le \sum_{j=0}^{l-1} \min\{a[i][j], b[i][j]\} \le |A' \cap B'| + \varepsilon(|A| + |B|)$$

$$|A' \cup B'| \le \sum_{j=0}^{l-1} \max\{a[i][j], b[i][j]\} \le |A' \cup B'| + \varepsilon(|A| + |B|)$$

Therefore, according to Lemma 3.4,

$$\hat{J}_{AB}^{(i)} \le \frac{|A' \cap B'|}{|A' \cup B'|} + \frac{\varepsilon(|A| + |B|)}{|A' \cup B'|} \le J_{A'B'} + \frac{\varepsilon}{1-\varepsilon} \cdot \frac{(|A'| + |B'|)}{|A' \cup B'|}$$

$$\le J_{AB} + 2\varepsilon + \frac{2\varepsilon}{1-\varepsilon}$$

On the other hand,

$$\hat{J}_{AB}^{(i)} \ge \frac{|A' \cap B'|}{|A' \cup B'| + \varepsilon(|A| + |B|)} \ge \frac{|A' \cap B'|}{|A' \cup B'|} \cdot \frac{1}{1 + \frac{\varepsilon}{1-\varepsilon} \cdot \frac{|A'|+|B'|}{|A' \cup B'|}}$$

$$= J_{A'B'} \cdot \frac{1}{1 + \frac{\varepsilon}{1-\varepsilon} \cdot (1 + J_{A'B'})} \ge J_{A'B'} \cdot \frac{1}{1 + \frac{2\varepsilon}{1-\varepsilon}}$$

$$\ge J_{A'B'} - \frac{2\varepsilon}{1+\varepsilon} \ge J_{AB} - 2\varepsilon - \frac{2\varepsilon}{1-\varepsilon}$$

which gives the proof. □

### 3.2 Simple version #2: The Count Version

Since the similarity is determined by the significant items, the remaining insignificant items only cause errors in the previous CM version. Notice that in the CM version, when the insignificant items collide with significant items, they will only increase the counters, which makes their effect add-up when there are multiple insignificant items collide into the same counter. While in the count sketch, the incoming item will increase or decrease the counter by 1 based on its hash value (as we introduced in 2.3.2), which will reduce other's effect when multiple insignificant items collide into a counter since some item may increase it by one while others decrease it by 1. Based on this idea, we come up with this count version using count sketch to estimate similarity.

Specifically, for multisets $A$ and $B$, we need two sets of $k$ counter arrays with length $l$, denote these counters by $a[i][j]$ and $b[i][j]$, where $i = 0, 1, ..., k-1$ and $j = 0, 1, 2, ..., l-1$. Also, there are $k$ independent hash functions $h_0, h_1, ..., h_{k-1}$ mapping items in $A \cup B$ randomly and uniformly into $\{0, 1, 2, ..., l-1\}$, and $k$ independent hash functions $s_0, s_1, ..., s_{k-1}$ mapping items in $A \cup B$ randomly and uniformly into $\{+1, -1\}$. For each coming item $x \in A$ and $y \in B$, the update process is:

$$a[i][h_i(x)] = a[i][h_i(x)] + s_i(x)$$
$$b[i][h_i(y)] = b[i][h_i(y)] + s_i(x)$$

After inserting all the items, we estimate $J_{A_{ij}B_{ij}}$ by:

$$\hat{J}_{A_{ij}B_{ij}} = \begin{cases} \frac{\min\{|a[i][j]|, |b[i][j]|\}}{\max\{|a[i][j]|, |b[i][j]|\}}, & \text{if } a[i][j] \cdot b[i][j] > 0. \\ 0,, & \text{otherwise.} \end{cases}$$

The idea is that intuitively, if there's an significant item mapped into $A_{ij}$ and $B_{ij}$, $a[i][j]$ and $b[i][j]$ will have the same sign. Therefore in this case we just do what we did as in the CM version to estimate $J_{A_{ij}B_{ij}}$. On the other hand, if $a[i][j]$ and $b[i][j]$ have different sign, we can see that $A_{ij}$ and $B_{ij}$ are not similar at all, therefore we'd simply estimate their similarity as zero.

As in equation 2, $J_{AB}$ can be written by a linear combination of $J_{A_{ij}B_{ij}}$. But we can not estimate $|A \cup B|$ and $|A_{ij} \cup B_{ij}|$ in the coefficient, so we simply take the average (i.e. simple combination where each $J_{A_{ij}B_{ij}}$ has the same weight) to give the estimation. Therefore, our estimator is:

$$\hat{J}_{AB} = \frac{1}{kl} \sum_{i=0}^{k-1} \sum_{j=0}^{l-1} \frac{\min\{|a[i][j]|, |b[i][j]|\}}{\max\{|a[i][j]|, |b[i][j]|\}} \cdot \mathbb{1}\{a[i][j] \cdot b[i][j] > 0\}$$

Detailed properties, results and analysis of the count version can be found in section 4.2. We can see that the count version is much more accurate than the CM version when using a small amount of memory.

### 3.3 Formal version

As we see in the count version, using the count sketch can do better in estimating the subset similarity, but it can not estimate the combination coefficients well. While the CM version can estimate these coefficients pretty well, so intuitively we'd combine the CM version and the count version together, using the count version to estimate the subset similarity and using the CM version to estimate the combination coefficients, which leads to our formal design of SimiSketch.

Formally, SimiSketch works in the following way. for multisets $A$ and $B$, we need two sets of $k$ counter arrays with length $l$, denote these counters by $a[i][j]$ and $b[i][j]$, where $i = 0, 1, ..., k-1$ and $j = 0, 1, 2, ..., l-1$. Each counter slot has two fields, namely cm and c. Also, there are $k$ independent hash functions $h_0, h_1, ..., h_{k-1}$ mapping items in $A \cup B$ randomly and uniformly into $\{0, 1, 2, ..., l-1\}$, and $k$ independent hash functions $s_0, s_1, ..., s_{k-1}$ mapping items in $A \cup B$ randomly and uniformly into $\{+1, -1\}$. For each coming item $x \in A$ and $y \in B$, the update process is:

$$a[i][h_i(x)].cm = a[i][h_i(x)].cm + 1$$
$$a[i][h_i(x)].c = a[i][h_i(x)].c + s_i(x)$$
$$b[i][h_i(y)].cm = b[i][h_i(y)].cm + 1$$
$$b[i][h_i(y)].c = b[i][h_i(y)].c + s_i(x)$$

After inserting all the items, we use the c field to estimate the subset similarity, and use the cm field to estimate the combination coefficients in equation 2. After that, we take the average on the $k$ rows to give the estimation. Specifically, our estimator to $J_{AB}$ is given by

$$\hat{J}_{AB} = \frac{1}{k} \sum_{i=0}^{k-1} \sum_{j=0}^{l-1} \frac{\max\{a[i][j].cm, b[i][j].cm\}}{\sum_{k=0}^{l-1} \max\{a[i][k].cm, b[i][k].cm\}} \cdot \hat{J}_{A_{ij}B_{ij}}, \text{ where}$$

$$\hat{J}_{A_{ij}B_{ij}} = \begin{cases} \frac{\min\{|a[i][j].c|, |b[i][j].c|\}}{\max\{|a[i][j].c|, |b[i][j].c|\}}, & \text{if } a[i][j].c \cdot b[i][j].c > 0. \\ 0,, & \text{otherwise.} \end{cases}$$

### 3.4 Optimization using SALSA

SALSA (Self-Adjusting Lean Streaming Analytics) [7] is a practical technique to optimize sketching algorithms. As a brief introduction to SALSA, consider count-min sketch. Practically each counter

SimiSketch: Efficiently Estimating Similarity of streaming Multisets

in the count-min sketch takes up 4 bytes memory, since 4-byte counters don't overflow often. But in fact 4-byte is a waste for many counters in the count-min counter array, since their values are not that big, maybe 2-byte is enough for them. So the idea of SALSA is that allocate small size counters at first, for example only 1 byte for each counter. Then there will be much more counters compare to using 4-byte counters, which will help to reduce the conflicts and increase the accuracy. When a counter is about to overflow, we just combine it with its nearby counter to get a counter of bigger size to avoid overflow. So the SALSA technique can help use memory more efficiently, as the result create more available counters, reduce conflicts of different items and increase the accuracy.

We can also use SALSA to optimize SimiSketch. Specifically, initially in each counter the cm and c field both have 1 byte. There's also an 1-bit indicator for each counter indicating whether it's combined with its neighbors and which counter it belongs to. Logically we consider each counter array as a ring, and each time when a counter overflows we combine it with its neighbor in the clockwise direction. Each time when two counters combine, we add their cm field together, and the length of the cm field in the new counter equals to the sum of that of the two previous cm fields, and so as the c field. Also we flip the 1-bit indicator to show that they are combined into a bigger counter. Notice that the hash functions $h_i$ maps items into initial tiny counters (1-byte counters), and they remain unchanged after combination occurs. After the combination, when an item is mapped into an 1-byte counter which is a part of a larger counter, it will add into the larger counter. After inserting all the items in multisets $A$ and $B$, notice that their counters may not be aligned (since their counts may combined in different ways during insertion), so we need to do some further combination to make them aligned. Then we can use the same way stated in section 3.3 to estimate $J_{AB}$.

# 4 EVALUATION

## 4.1 Setup

**Datasets:** To perform evaluation, we use three different datasets as below.

- **Synthetic Dataset:** the synthetic dataset is generated according to the Zipf distribution. We vary the skewness from 0.1 to 1.0. Each synthetic dataset contains about 200K distinct items. We split the dataset randomly into two subsets, and estimate the similarity of them.
- **IP Trace Dataset:** the IP trace dataset contains one hour of anonymous network traces collected from the Equinix-Chicago monitor in 2018 [2]. We use the source IP and destination IP as the ID of items. We divide it into 1-minute intervals, and each contains around 27M items and 85K distinct items. We use data from two different intervals as two subsets, and estimate the similarity of them.
- **Text Dataset:** the text dataset contains article published in New York Times [23]. In order to get a larger multiset, we concatenate multiple articles together. Each multiset contains 0.6M items and 0.8M distinct items.

**Baselines:** We compare our algorithm with the state-of-the-art methods introduced in section 2.2. Since all of them can only deal with set, we use either a count-min sketch or a hash table to convert the multiset into a set (as stated in 1.1), and then feed them to these algorithms. Their parameter settings listed below.

- **MinHash:** we take $k = 128$ (as in 2.2.1).
- **HyperLogLog:** we take $N = 64$ and $M = 11$ (as in 2.2.2).
- **MaxLogHash:** we take $k = 128$ (as in 2.2.3).
- **DotHash:** since our dataset is pretty large, we tried up to $d = 10000$ (as in 2.2.4), ended up to a very bad result and very low throughput. The bad result is caused by the fact that $d \ll |A|, |B|$, so that the approximate orthogonality isn't hold anymore in this case. The low throughput is caused by heavy hash function computation, since we need to call about $d = 10000$ hash functions each item occurs. So we decide not to plot DotHash in our figures.

**Metrics:** Our comparison with baselines focuses on accuracy and throughput. For accuracy we use metric **RE (Relative Error)**, defined as $RE := \frac{\hat{J}_{AB} - J_{AB}}{J_{AB}}$. For throughput we use metric **MIPS (Million Items Per Second)**.

**Implementation:** We implemented SimiSketch and all the baseline algorithms in C++, and run on a server with dual 18-core CPUs (36 threads, Intel Core i9-10980XE CPU @3.00GHz) and 125GB DRAM memory. For hash functions, we use Farmhash [1] in our implementation. Our source code is open-sourced on GitHub for reference [3].

## 4.2 Test on Parameter Settings

In this section, we conduct tests on the parameter settings of our algorithm. Specifically, we test on selection of $k$ (number of rows), allocated memory and flow distribution separately. We use the IP trace dataset to conduct tests on different number of rows ($k$) and memory allocation. In order to test the effect of different data distribution, we use synthetic dataset with different Zipf parameter.

**Number of rows ($k$) (figure 2):** We tested the selection of $k$ using 500KB and 2MB memory, the results are shown in figure 2(a) and 2(b) respectively. In both figure we can see that for the CM version, when we increase $k$ its error will increase. For the count version, when we increase $k$ its error will decrease. For the formal version, its error almost remains the same as we increase $k$. Notice that as we fix the size of memory, as we increase $k$, the length of each array ($l$) will decrease. So in the case of CM version, there will be significantly more conflicts as we increasing $k$, leading to the increase of error. As for the count version, we find that it may perform better with a small amount of memory (as shown below), which explains its increase of accuracy as $k$ increases.

**Memory (figure 3):** We tested the accuracy using different size of memory varying from 10KB to 2MB. For each version of SimiSketch we pick $k = 1$. As we see in figure 3, when the memory consumption increase, the accuracy of both CM version and the formal version increases. However, the error of count version decreases at fist (3.35% at 10KB to 0.37% at 40KB), then begins to increase (till −46.7% at 2MB). This is caused by the fact that the estimation given by the count version is not coherent. Specifically, it's easy to see that if we use lots of memory so that there's no conflicts at all, the CM version and the formal version will give the precise estimation without any error, but the count version won't. Since we can not estimate the combination coefficients in equation 2 in the count version, we simply take the average over all

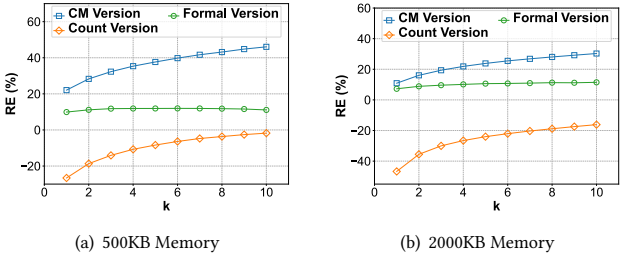

(a) 500KB Memory  (b) 2000KB Memory

Figure 2: Experiments on different number of rows ($k$).

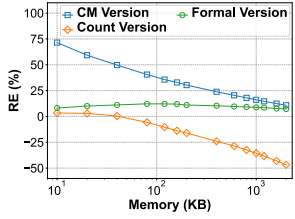

Figure 3: Experiments on different memory size.

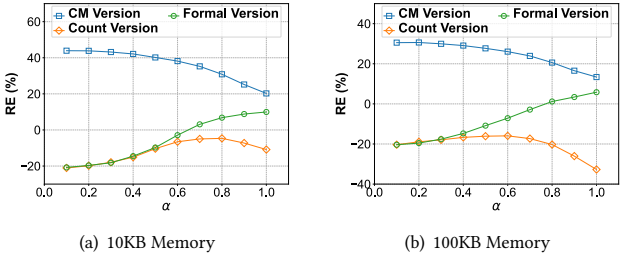

(a) 10KB Memory  (b) 100KB Memory

Figure 4: Experiments on different flow distribution.

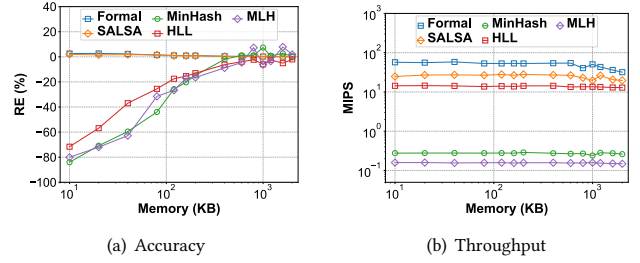

(a) Accuracy  (b) Throughput

Figure 5: Text dataset, with count-min sketch.

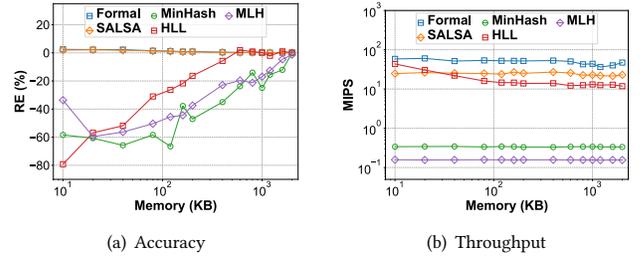

(a) Accuracy  (b) Throughput

Figure 6: Text dataset, with hash table.

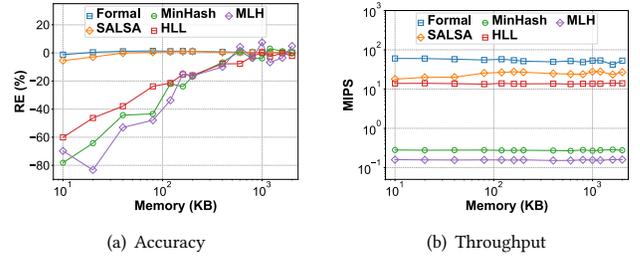

(a) Accuracy  (b) Throughput

Figure 7: IP trace dataset, with count-min sketch.

the subsets. Since the number of significant items isn't too much, we can estimate their similarity $J_{A'B'}$ (as in 3.1) using pretty small amount of memory. As we increase the memory, it will create more subsets containing only insignificant items, which contributes little to $J_{AB}$. Therefore, taking average over all these sets will decrease the accuracy.

Even though the count version can provide really good estimation with a small amount of memory, it's in fact not that practice since it's hard to determine it's ideal memory size. If we give it more or less, it will end up in a poor accuracy. High level speaking, the CM version is good at large memory but bad at small memory, the count version is good at small memory but bad at large memory. As for our formal version, it's as good as the count version at small memory, and as good as the CM version at large memory, which makes it really an ideal and practical solution.

**Flow distribution (figure 4):** We test the effect of different flow distributions using data generated from Zipf distribution, its skewness parameter $\alpha$ varying from 0.1 to 1.0. We test on all of the three versions taking $k = 1$, using 10KB and 100KB memory, the result is in figure 4(a) and 4(b) respectively. We see that our algorithm is doing well in all these distributions, and especially good at dealing with those which are more skew (with larger $\alpha$). Since in this case the significant items will be less, and from our previous analysis in 3.1-3.3, our algorithm will do better.

### 4.3 Comparison with Baselines

We compare our formal version with/without SALSA optimization with other baseline algorithms on accuracy and throughput. We use two real-world datasets, the text dataset and the IP trace dataset. The result is shown in figure 5-8. In the legends, "formal" represents our formal version without SALSA, "SALSA" represents our formal version with SALSA, "HLL" represents HyperLogLog, "MLH" represents MaxLogHash. We can see that our algorithm is of higher accuracy and higher throughput than any other baseline in any memory settings.

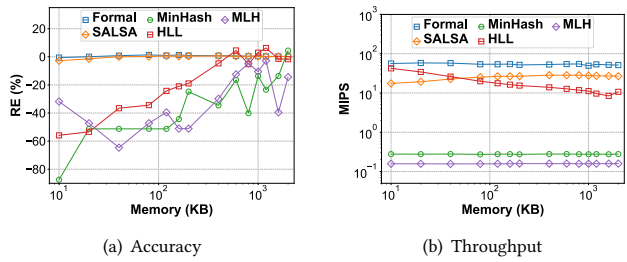

(a) Accuracy  (b) Throughput

Figure 8: IP trace dataset, with hash table.

## 5 CONCLUSION

In this paper, we propose the SimiSketch, solving the classical set similarity estimation problem in a new way. As our best knowledge, we are the first one using sketching algorithms to estimate the set similarity. We came up with several versions of SimiSketch, from the more intuitive but less practical CM and count versions, to the accurate and practical but more complicated formal version and the SALSA-optimized version. We run tests on real-world networking and text dataset, showing that compared to the state-of-the-art solutions, our algorithms improve the accuracy by up to 42 times, and increase the throughput by up to 360 times.

## ACKNOWLEDGMENTS

We would like to thank the anonymous reviewers for their valuable suggestions. This work is supported by Key-Area Research and Development Program of Guangdong Province 2020B0101390001, and National Natural Science Foundation of China (NSFC) (No. U20A20179).